\documentclass[]{raa} 

\usepackage{graphicx,times}             

\begin{document}

   \title{Network and Plage Indices from Kodaikanal Ca-K Data
}

   \volnopage{Vol.0 (200x) No.0, 000--000}      
   \setcounter{page}{1}          

   \author{K. P. Raju
      \inst{}
   \and Jagdev Singh
      \inst{}
   }

   \institute{Indian Institute of Astrophysics
             Bangalore, India ; {\it kpr@iiap.res.in}
   }

   \date{Received~~2009 month day; accepted~~2009~~month day}

\abstract{The Ca II K filtergrams from Kodaikanal Solar Observatory have been 
used to study the solar activity. The images are dominated by the chromospheric 
network and plages. Programs have been developed to obtain the network and 
plage indices from the daily images as functions of solar latitude and time. 
Preliminary results from the analysis are reported here. The network and plage 
indices were found to follow the sunspot cycle. A secondary peak is found 
during the declining activity period in both the indices. A comparison of 
network indices from the northern and the southern hemispheres shows that the 
former is more active than latter. However such an asymmetry is not clearly seen 
in the case of plage index.
}

   \authorrunning{K. P. Raju \& J. Singh }            
   \titlerunning{Solar Activity Indices from Kodaikanal }  

   \maketitle

\section{Introduction}           
\label{sect:intro}

The Calcium II K images of the Sun from Kodaikanal have a data span of about 
100 years and contain information of more than 9 solar cycles. This provides a good 
opportunity to study the solar activity in great detail. Only Mt. Wilson in USA 
has a comparable data span (Foukal et al. 2009). The Ca II K line  at 3934 {\AA} is a 
chromospheric emission line and the images are dominated by the chromospheric 
network and plages. The chromospheric network is the bright emission network 
which outline the supergranulation cells (Simon and Leighton 1964). Convective 
motions in the 
supergranules sweep magnetic field elements - small flux tubes - to their edges,
resulting in field concentrations which produces enhanced emission at the cell 
boundaries. The plage is the bright chromospheric region, part of active region outside 
sunspots.

The supergranular convection plays an important role in maintaining the solar cycle.
When the active regions on the solar surface disintegrate, magnetic flux disperses along 
supergranular boundaries through a random walk process and neutralizes the polar 
fields. As the Ca K chromospheric network is a manifestation of supergranulation 
phenomenon, the Ca line is a good tracer of magnetic activity. Although 
total solar irradiance variations are only of the order 0.1 \% over solar cycle, the UV 
irradiance show much larger variations (Johannesson, Marquette, \& Zirin 1998).
The UV radiation in the upper atmosphere is believed to play a role in the Earth's 
climate. The Ca K line is a proxy to the UV irradiance and hence it is particularly 
useful in the pre-satellite era.

In this preliminary study, we have used about 100 Ca K filtergrams, one from every 
month, during the period 1997-2007, from Kodaikanal to obtain the activity indices. 
The indices represent fractional area of the feature over the solar disk.
We have developed codes for calculating network and plage indices. The data analysis 
and some preliminary results are explained in the following sections.

\section{Observations}
\label{sect:Obs}

The routine Ca K observations have been going on in Kodaiakanal since 1906 (Bappu 1967).
The spectroheliograms which were obtained in photographic plates are presently being 
digitized and calibrated. In 1995, the observational setup was changed to get filtergrams 
using a narrow-band (pass band = 0.12 nm) Ca K filter and a CCD camera with 1Kx1K 
format, but with the same old siderostat. In the present work, these filtergrams are used.
It may be noted that there are gaps in the data due to the bad weather and instrument problems. 
The long gaps in the data are mainly seen in the rainy season.
A typical example of the filtergrams is shown in Figure 1. The pixel resolution is about 
2.5 arc sec. This arrangement was replaced by a twin telescope in 2008 (Singh et al. 2012).

\section{Data analysis}
\label{sect:analysis}

The Ca K filtergram is usually taken from the middle of the month. When 
this is not avaialble, the one from a nearby day is taken. The data analysis 
involved the following steps.

1. Correction for image rotation: Since a siderostat was used in the observations,
the images rotate with time. Hence the filtergrams need to be corrected for the rotation.

2. Centre and radius of the solar image: The (X,Y) positions of the limb were noted
and then fitted with a circle which gives the center and radius of the image.

3. Background intensity: The quiet-Sun background intensity was determined according 
to a technique described by Brandt \& Steinegger (1998). The solar disk is divided 
into 50 concentric rings of equal area (Figure 1). In each of the ring, the 5 \% value
in the cumulative intensity histogram is found and all intensity values in the ring 
are replaced by this value. The quiet-Sun background intensity thus determined was 
used to normalize the individual images. This will also correct for the limb darkening. 

   \begin{figure}
   \centering
   \includegraphics[width=\textwidth, angle=0]{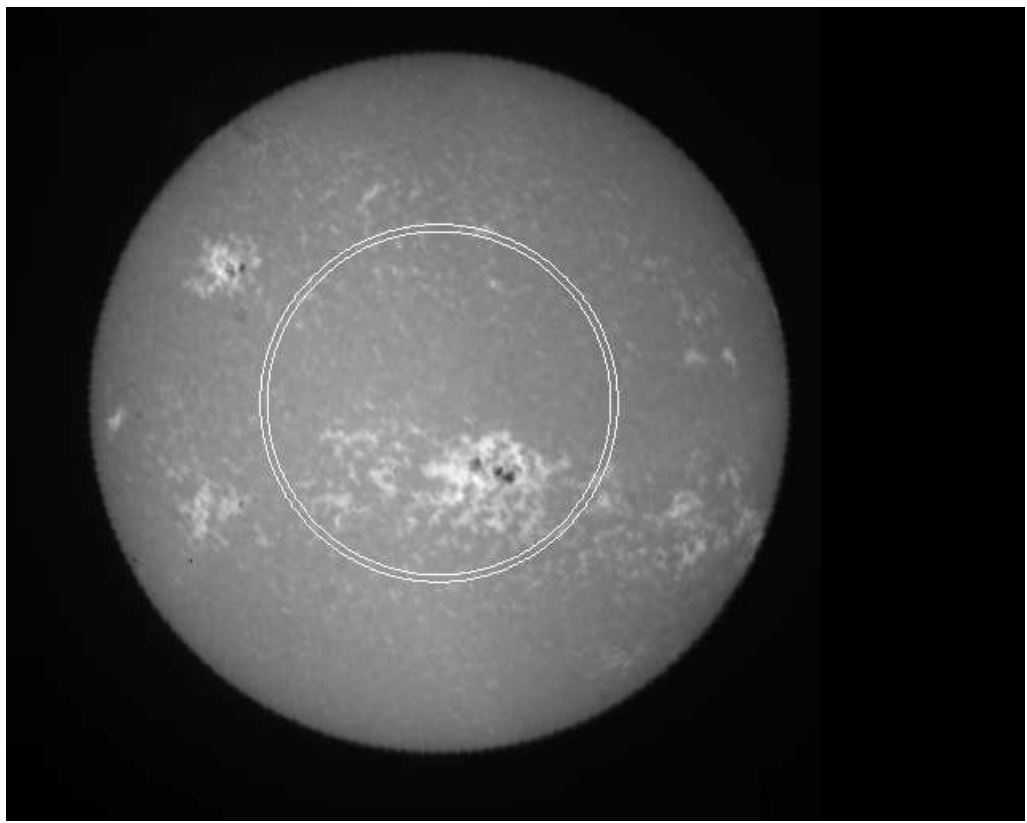}
   \caption{The Ca K image from Kodaikanal observatory. One of the 50 concentric rings 
     of equal area, used in finding the background intensity is also shown.}
   \label{Fig1}
   \end{figure}

4. Polynomial fit: Each 10 degree latitude strips of the normalized image were fitted 
with a second degree polynomial surface.  The strip was then divided by the polynomial
fit which will smoothen the sharp gradients. 

5. Identifying network \& plage points: Now the intensity thresholds for network and 
plage were decided by trial and error. They were found to be 3 \% and  10 \% respectively.

\section{Results and Discussion}
\label{sect:discussion}

The network and plage indices were obtained from the images. The variation of network 
index during the period 1997-2007 is given in Figure 2. The monthly sunspot number is 
also plotted alongwith. It may be seen that the network index generally follows the 
sunspot cycle. 
A secondary peak can also be seen during the period of declining activity. The network
index shows a variation of about 8 \% between the solar maximum and minimum.

   \begin{figure}
   \centering
   \includegraphics[width=\textwidth, angle=0]{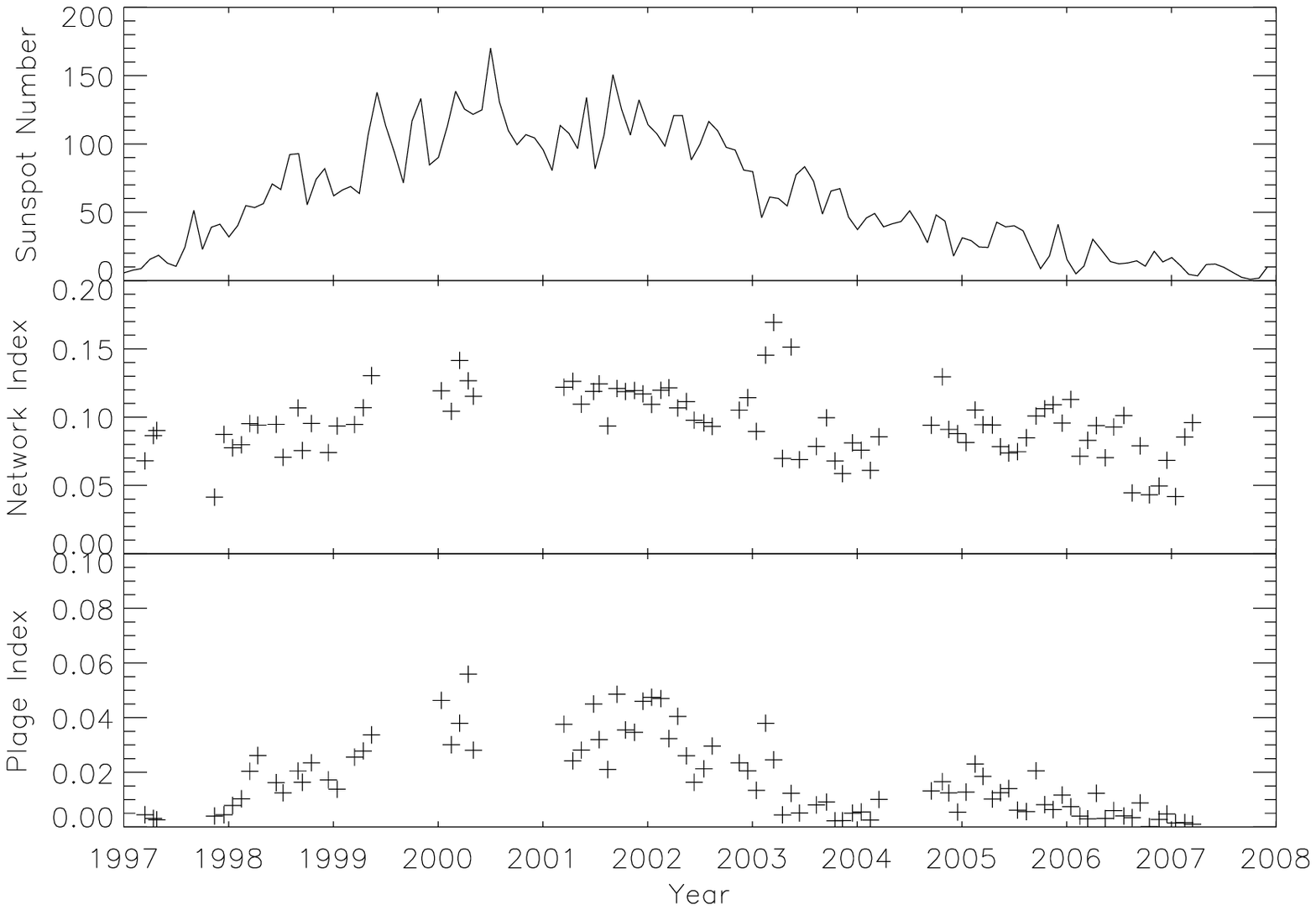}
   \caption{Variation of network and plage indices with time. The 
     monthly sunspot number is plotted as solid line in the top panel.}
   \label{Fig2}
   \end{figure}

In Figure 2, variation of plage index is plotted for the same period. The behaviour of 
plage index is similar to that of the network index. The plage index follows the solar 
cycle more clearly than the network index and also shows evidence of the secondary peak.
The variation between the solar maximum and minimum is about 5 \%.

The value of  network index found by us (0.05 -- 0.13) is much less than the value 
usually reported which is up to 0.3 (Worden et al. 1998, Singh et al. 2012). This 
is because of the higher thresholds chosen by us, in order to avoid the doubtful 
network points. Hence our values can be taken as the enhanced network indices. 
The reason of the origin of the secondary peaks is not clear but could be due the 
emergence of excess active regions in the declining period.

   \begin{figure}
   \centering
   \includegraphics[width=\textwidth, angle=0]{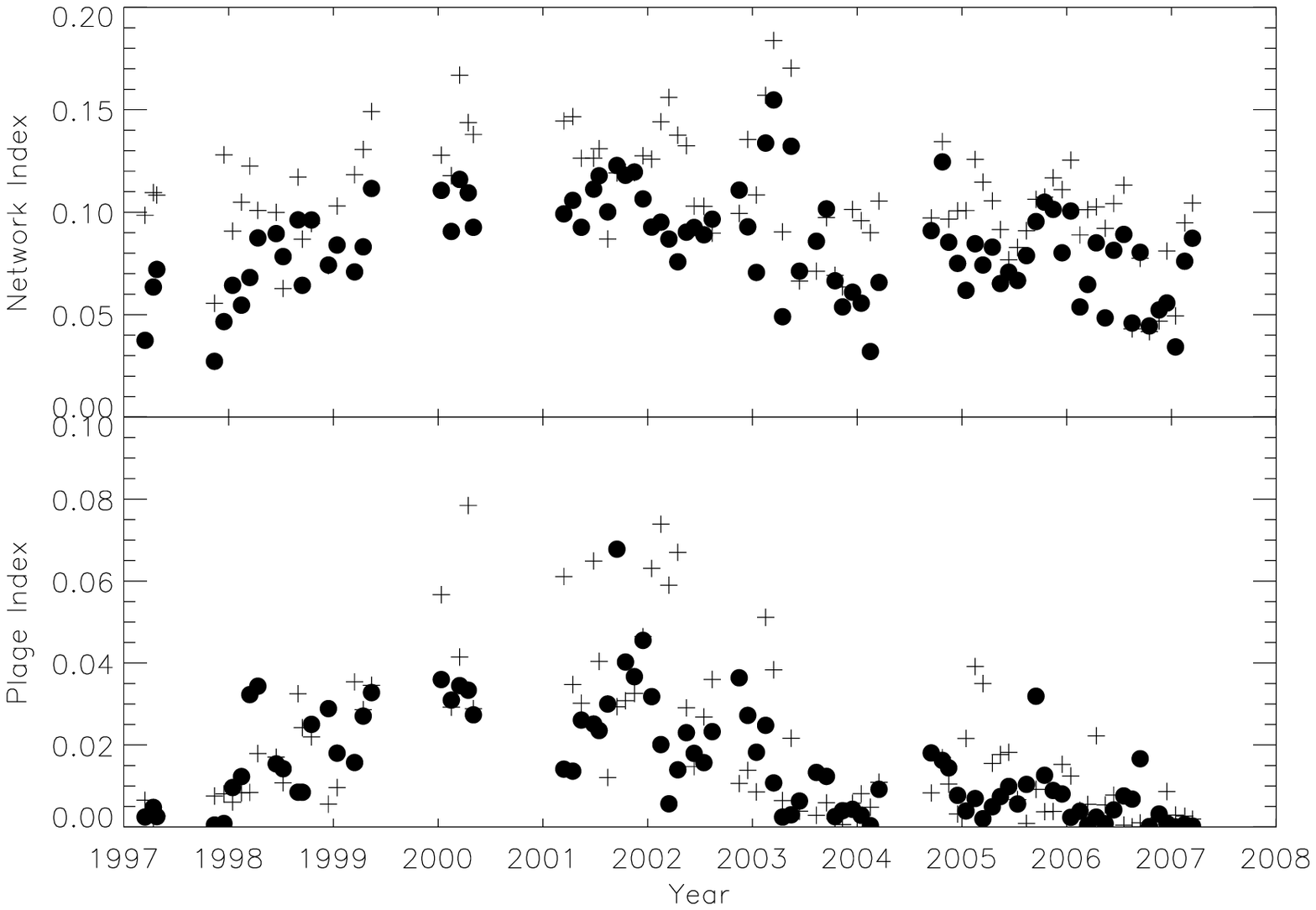}
   \caption{North-South differences in network and plage indices. The 'plus' indicates 
points from the northern hemisphere and filled circle indicates those from the 
southern hemisphere.}
   \label{Fig3}
   \end{figure}

The differences in the north and south hemispheres of the indices were also obtained. 
These are plotted in Figures 3. It can be seen that in the case of the network index, 
northern hemisphere is more active than the southern hemisphere. However, the plage 
index does not show such a clear difference between the north and south hemispheres.

The north-south asymmetry in the activity in the solar cycle 23 has been reported 
(Joshi and Joshi 2004, Li et al. 2009, Bankoti et al. 2010). An overall dominant 
southern hemisphere is observed. The northern hemisphere is more active during the 
ascending years of the cycle but there is a transition to more active southern 
hemisphere on the later years. Our results, in general, do not agree with the above 
findings, although there is some consensus regarding the activity during the 
ascending years. This aspect needs to be examined in more detail.

\section{Conclusions}
\label{sect:conclusion}

A preliminary study of solar activity using Ca K filtergrams from Kodaikanal during 
the period 1997-2007, which mostly covers the solar cycle 23, has been carried out. 
The network and plage indices were obtained during the period and they were found to 
follow the sunspot cycle.  A secondary peak during the declining activity was found in 
both network and plage indices. A comparison of network indices from the northern and the 
southern hemispheres shows that the former is more active than latter. However such an 
asymmetry is not clearly seen in the case of plage index.

As a future work, the activity indices will be obtained from the 100 year data. The 
behavior of the indices in the individual latitude bands will also be be studied.

\begin{acknowledgements}
This work was funded by the Department of Science and Technology, Government of India.
\end{acknowledgements}

\label{lastpage}

\end{document}